\documentclass[10pt,letterpaper]{article}
\usepackage{opex3}
\usepackage{cite}
\usepackage{amsmath,amssymb,graphicx}
\begin{document}
\title{Ultrafast, low-power, all-optical switching via birefringent phase-matched transverse mode conversion in integrated waveguides}
\author{Tim Hellwig,$^{*,1}$ J\"orn P. Epping,$^{2}$ Martin Schnack,$^{1}$ Klaus.-J. Boller,$^{2}$ and Carsten Fallnich$^{1,2}$}

\address{$^{1}$Institute of Applied Physics, Westf\"alische Wilhelms-Universit\"at, Corrensstrasse 2,\\ 48149 M\"unster, Germany
\\$^{2}$Laser Physics \& Nonlinear Optics Group, MESA+ Institute for Nanotechnology,\\ University of Twente, P. O. Box 217, Enschede 7500AE, The Netherlands}

\email{$^{*}$Corresponding author: tim.hellwig@uni-muenster.de}

\begin{abstract}
We demonstrate the potential of birefringence-based, all-optical, ultrafast conversion between the transverse modes in integrated optical waveguides by modelling the conversion process by numerically solving the multi-mode coupled nonlinear Schroedinger equations. The observed conversion is induced by a control beam and due to the Kerr effect, resulting in a transient index grating which coherently scatters probe light from one transverse waveguide mode into another. We introduce birefringent phase matching to enable efficient all-optically induced mode conversion at different wavelengths of the control and probe beam. It is shown that tailoring the waveguide geometry can be exploited to explicitly minimize intermodal group delay as well as to maximize the nonlinear coefficient, under
the constraint of a phase matching condition. 
The waveguide geometries investigated here, allow for mode conversion with over two orders of magnitude reduced control pulse energy compared to previous schemes and thereby promise nonlinear mode switching exceeding efficiencies of 90\% at switching energies below 1\,nJ.
 \end{abstract}

\ocis{(130.4310) Integrated optics, nonlinear; (190.4390) Nonlinear optics, integrated optics; (190.3270) Nonlinear optics, Kerr effect; (190.4420) Nonlinear optics, transverse effects in; (130.2755) Integrated optics, glass waveguides; (130.4815) Integrated optics, optical switching devices.}

\section{Introduction}
 
Efficient all-optical switching based on a number of nonlinear effects is a research topic of high interest \cite{Boyd1992}. In recent years all-optical switching at low pump powers in integrated optical circuits has been a particular focus of research, pushing the switching energies from the picojoule \cite{Almeida2004} into and even below the femtojoule regime \cite{Tanabe2007,Hu2008a,Nozaki2010}. These advances were made possible by switching light in high quality-factor nanocavities by modifying the effective index of the cavity using, e.g., pump-induced free-carrier injection and thereby shifting the resonant frequency of the cavity. Using resonant field enhancement, however, comes at the cost of narrowing the usable optical bandwidth and demands precisely tunable lasers or some means to tune the cavity resonances \cite{Wild2004}. A broader bandwidth is offered by all-optical switching in non-resonant linear geometries, which has been achieved by, e.g., nonlinear-phase shifts in Mach-Zehnder type interferometers \cite{Nakamura2004} or by nonlinear frequency conversion combined with frequency filtering \cite{Koos2009}, which has recently been demonstrated in a mode-selective manner \cite{Ding2014}. As an alternative, transient long period gratings, optically induced by multi-mode interference in combination with the nonlinear Kerr-effect, look promising for the realization of a low-power, all-optical device for switching in a linear geometry. These long-period gratings can be used to convert the transverse modal content of a probe beam. Spatial conversion or switching is of special interest in the context of future spatial-division multiplexing networks in addition to or instead of all-optical frequency conversion \cite{Richardson2013}. All-optical mode conversion based on optically induced long-period gratings (OLPG) has been demonstrated  experimentally in a few-mode fiber using nanosecond pulses with 70 $\mu$J pulse energy \cite{Andermahr2010}. Recently, femtosecond control beams were used to induce the long-period gratings to allow for high peak powers at substantially reduced pulse energies \cite{Hellwig2014a}. A severe restriction of the used setup for applications is, however, that the pulse energies are still high, in the order of 100\,nJ, which requires high-power amplifier systems. Additionally, nonlinear-polarization rotation was found to introduce a phase-sensitive cross-talk between the cross-polarized control and probe beam, resulting in a demanding probe separation.

Here, we show low-power, cross-talk free all-optical conversion of transverse modes in integrated waveguides by numerically modeling the nonlinear interaction between the involved pump and probe beam modes. The key to such progress lies in the use of integrated waveguides in which one can make use of a highly nonlinear material, e.g., silicon nitride ($\text{Si}_3\text{N}_4$), featuring a nonlinear index of refraction ($n_2=2.4\cdot10^{-15}\,\text{cm}^2/\text{W}$) that is ten times larger than in fused silica optical fibers. The higher non-linearity in combination with modal areas being three orders of magnitude smaller in waveguides compared to multimode fibers, allows to accordingly lower the required peak powers. The aforementioned cross-talk induced by nonlinear polarization rotation can be avoided if the control and probe beam could be separated by other means than polarization. A dichroic setup with different control and probe wavelengths would be a favorable approach but requires additional means for phase matching, i.e., a matching of the induced Kerr grating to the phase velocities of the probe beam. We show that this can be achieved by tailoring the waveguide cross section to induce an appropriate birefringence that provides phase matching for selectable, well separated control and probe beam wavelength combinations (for details on the phase matching see section~\ref{sec:pm}). Thereby, a cross-talk free operation by separating control and probe beam by wavelength is indeed possible.
Compared to earlier work \cite{Andermahr2010,Hellwig2014a,Hellwig2013} the small modal areas inherent to integrated waveguides are coming at the cost of an increased group velocity mismatch of the involved modes of up to two orders of magnitude. However, this is more than compensated by a much higher effective nonlinearity, which substantially reduces the necessary pulse energy for all-optical mode conversion.
An important advantage of our approach to use form-induced waveguide birefringence is that it can be applied to other materials as well, e.g., to silicon or chalcogenide glasses, with an $n_2$ about a factor of 300 - 500 times higher than in fused silica. In the calculations presented here, we have selected  $\text{Si}_3\text{N}_4$, due to its availability with loss as low as 0.1\,dB/m \cite{Bauters2011} and a transparent wavelength range from 400\,nm to about 2.3$\,\mu$m, making $\text{Si}_3\text{N}_4$ an experimentally widely used material for nonlinear optics \cite{Ikeda2008,Levy2009a,Levy2011a,Foster2011,Epping2013}. 

\section{Mode conversion scheme}
\begin{figure}[htb]
\centering
\includegraphics[width=1\textwidth]{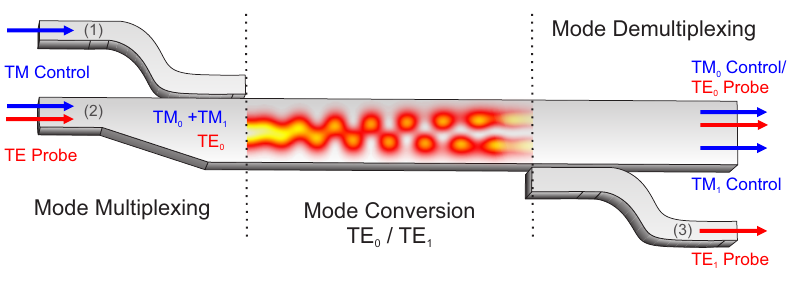}
\caption{Schematic of the proposed all-optical mode conversion scheme (dimensions not to scale). Potentially, tapered directional couplers could be used to excite a superposition of the $\text{TM}_0$- and $\text{TM}_1$-mode at the control wavelength as well as the $\text{TE}_0$-mode at the probe wavelength in the multimode waveguide channel marked by 2. In this work, only the nonlinear interaction of control and probe beam in the straight section of the multi-mode waveguide and the corresponding mode conversion to the higher-order mode is studied by numerically solving the multi-mode coupled nonlinear Schroedinger equations. An illustration of the probe mode's intensity profile along the propagation path is shown color-coded in the waveguide to depict the conversion process. For details see text.}
\label{setup_fig}
\end{figure}

The scheme for all-optical mode conversion that we model is depicted in Fig.~\ref{setup_fig}. The size and material parameters are chosen to enable an experimental verification in a subsequent step and take into consideration recently available $\text{Si}_3\text{N}_4$ waveguides with thick cross-sections \cite{Epping2014b}. The scheme consists of three parts: the excitation of the specific control and probe modes by multiplexing via two input waveguide channels (labeled 1 and 2), the actual all-optical mode conversion in the waveguide channel 2, and finally a mode demultiplexing into an output channel (labeled 3).

The control and probe beams are launched into cross-polarized modes for making use of the waveguide birefringence in order to enable phase matching and thus to achieve high conversion efficiency at different control and probe wavelengths (for details on the phase matching see section~\ref{sec:pm}). In case of the rectangular waveguide geometry investigated here the cross-polarized quasi-transverse electric (quasi-TE) and quasi-transverse magnetic (quasi-TM) modes are used.
In order to excite an OLPG with maximum contrast, the control beam energy is equally distributed between two different modes. We consider to use a multiplexing scheme based on tapered directional couplers \cite{Ding2013} to excite a specific mode mixture in the multimode waveguide section. A control beam at a center wavelength $\lambda_\text{c}$ is then coupled with equal powers into the two single-mode waveguide channels 1 and 2 (see Fig.~\ref{setup_fig}). During its propagation in the tapered section of channel 2 the coupled fundamental mode adiabatically evolves into the fundamental mode of the multimode section of the channel. If the propagation constants of the fundamental mode in channel 1 and that of the higher-order mode in channel 2 are matched, the light in channel 1 is coupled into the higher-order mode of the main waveguide channel 2 \cite{Ding2013}. The superposition of both modes within the straight mode conversion section of channel 2 then forms a transversely structured interference pattern which induces, via the Kerr effect, the OLPG.
When injecting a probe beam at a center wavelength of $\lambda_\text{p}$ into the main waveguide channel 2, it first adiabatically evolves into the fundamental mode. In the straight section of channel 2 the probe mode is then converted into the higher-order mode (schematically depicted in the middle section of Fig.~\ref{setup_fig}) by means of the OLPG. The amount of probe pulse energy that is converted by the OLPG is directed by demultiplexing in the adiabatic output coupler to output port 3, while otherwise the probe pulse would leave the straight output port unmodified.

In the following, we demonstrate the nonlinear conversion of the probe beam's modal content within the mode conversion section of the waveguide channel 2 (see Fig.~\ref{setup_fig}) by numerically solving the multi-mode coupled nonlinear Schroedinger equations \cite{Poletti2008}. The dispersion of the transverse waveguide modes is calculated up to the fifth order from the wavelength-dependent propagation constants obtained from using a vector finite difference mode solver \cite{Fallahkhair2008}. A power loss of $1\,\text{dB}/\text{cm}$ is assumed as a conservative estimate \cite{Epping2013}. During propagation in the straight section of channel 2 and conversion of the probe beam from the fundamental mode to the higher-order mode the power conversion efficiencies are analyzed. 

In section~\ref{sec:pm} the proposed birefringent phase matching scheme will be presented and it will be shown in section~\ref{sec:optim} that the waveguide dimensions can be tailored to achieve low-power mode-conversion within the boundary conditions set by the phase matching. Finally, in section~\ref{sec:conv} numerical simulations of all-optical mode conversion are presented for switching 3\,ps probe pulses utilizing 6\,ps long control pulses with the latter having peak powers ranging from 10\,W to 500\,W. The results of the performed simulations are compared for the different waveguide configurations discussed in section~\ref{sec:optim}. The potential applicability of our scheme for all-optical ultrafast switching with low pulse energies of less than a nanojoule is demonstrated.

\section{Phase matching}
\label{sec:pm}
The OLPG induced via the Kerr effect in waveguides arises from the multimode interference intensity pattern caused by the beating of two control beam modes (here either $\text{TE}_0$ and $\text{TE}_1$, or $\text{TM}_0$ and $\text{TM}_1$), that propagate with different phase velocities. In order to achieve efficient mode conversion the period of the induced OLPG  has to be adapted to the difference in propagation constants of the probe modes \cite{Bures2009}. One can therefore derive a phase matching condition depending on the propagation constants of the control as well as the probe modes:
	\begin{gather*}
		\Delta  = \Delta \beta_\text{c} -\Delta \beta_\text{p},\qquad \text{with}\\
 \Delta \beta_\text{c} = \beta_{0,1}(\lambda_\text{c})-\beta_{0,2}(\lambda_\text{c}), \qquad \Delta \beta_\text{p} = \beta_{0,3}(\lambda_\text{p})-\beta_{0,4}(\lambda_\text{p}),
	\end{gather*}
	with $\beta_{0,i}$ being the propagation constant of the fundamental mode and the higher-order mode of the control beam $(i=1,2)$ and of the probe beam $(i=3,4)$, respectively.
 
 Exact phase matching ($\Delta=0$) is in general and especially in radially symmetric waveguides not fulfilled for unequal control and probe beam wavelengths. The individual mode profiles expand at different rates into the cladding \cite{Ramachandran2005} when increasing their wavelengths. Therefore, the modes experience a different change in effective refractive index as a function of wavelength leading to a change in the difference of the propagation constants, $\Delta \beta$, and thereby resulting in a phase-mismatch ($\Delta\neq0$). Such a phase mismatch would prevent an efficient energy transfer between the probe modes.

\begin{figure}[htb]
\centering
\includegraphics[width=0.8\textwidth]{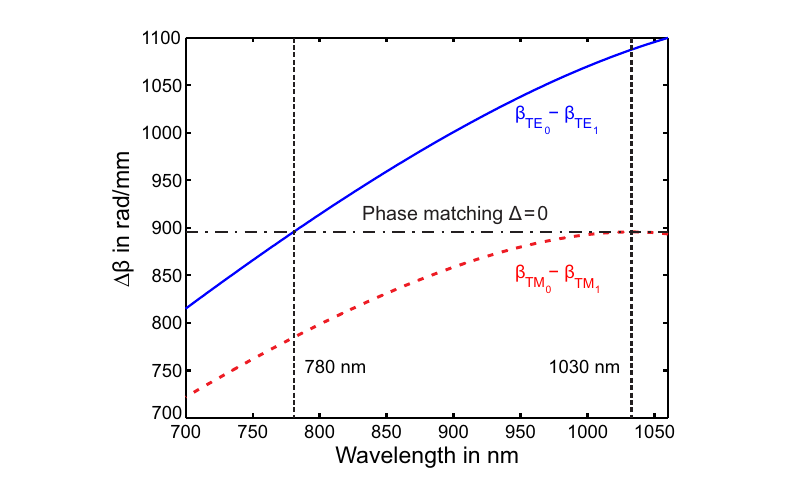}
\caption{Difference in propagation constants $\Delta \beta (\lambda)$ for TE- (blue curve) and TM-modes (red dashed curve) in a $\text{Si}_3\text{N}_4$ ridge waveguide in $\text{Si}\text{O}_2$ with an air on top cladding (ridge height: 380\,nm, ridge width: 949\,nm). The black dash-dotted line indicates  phase matching ($\Delta=0$) with a combination of 1030\,nm and 780\,nm for control and probe beam wavelengths.}
\label{fig:Phase_matching}
\end{figure}

Here we suggest a solution to this problem that is particular suitable for all-optical mode conversion in integrated waveguides. In rectangular waveguides, namely by choosing appropriately different transverse sizes, an inherent birefringence can be imposed and tailored to achieve phase matching at two different wavelengths. For an exemplary waveguide geometry the differences in propagation constants for the control and probe beam are depicted in Fig.~\ref{fig:Phase_matching} as the blue and dashed red curve, respectively. In this example, birefringent phase matching is fulfilled for a TM control beam at a center wavelength of 1030\,nm and a TE probe beam at about 780\,nm or vice versa. However, phase matching can also be achieved for a variety of wavelength combinations tailored specifically for the available external or integrated laser sources, e.g., also in the telecommunication window.  Here we focus on using the aforementioned wavelengths of 1030\,nm and 780\,nm as control and probe center wavelengths. For an experimental verification, both of these wavelengths can be easily generated by, e.g., an Yb-doped solid state or fiber laser and either a Ti:Sapphire-laser or a frequency-doubled Er-doped fiber laser. As these laser materials offer a relative wide gain bandwidth, fabrication tolerances can be compensated by a tuning of the laser frequency, although fabrication of $\text{Si}_3\text{N}_4$ waveguides has already been demonstrated with a standard deviation in the width of the waveguide of less than	 5\,nm \cite{Subramanian2013}.

\section{Waveguide optimization}
\label{sec:optim}
\begin{figure*}[htb]
\centering
\includegraphics[width=0.8\textwidth]{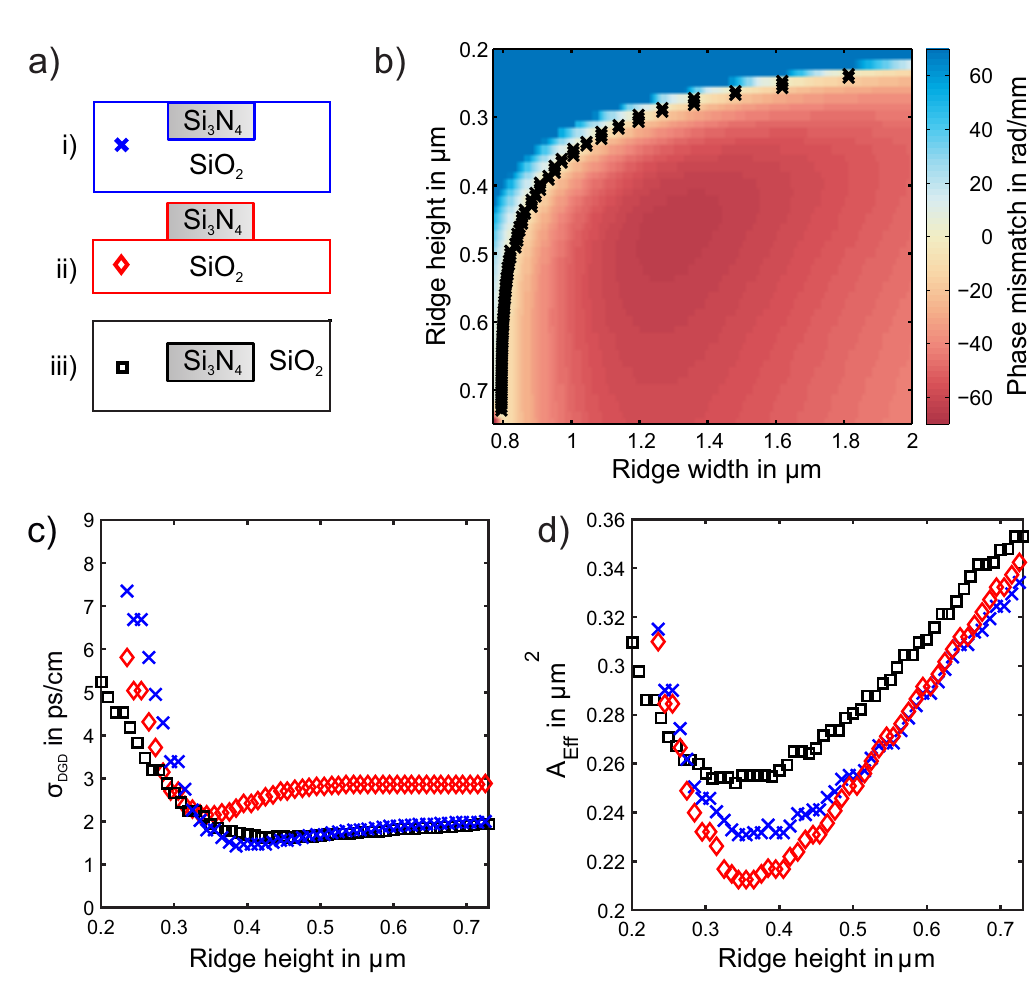}
\caption{(a) Schematic of the different cladding configurations of the rectangular waveguides discussed here. The symbols within the depicted cladding (black cross, red diamond and blue square) are the ones used in (c) and (d) to distinguish between the different cladding configurations. (b) Phase mismatch as a function of the waveguide dimensions for mode conversion at a center wavelength of 780\,nm via an optically induced long-period grating induced by a control beam at a center wavelength of 1030\,nm or vice versa. The cladding configuration assumed here is shown as (i) in (a). Waveguide dimensions that fulfill the phase matching condition are marked with black crosses. (c) Standard deviation of the group delays per unit length ($\sigma_{\text{GD}}$) of the involved modes for cladding configurations (i-iii) as a function of the ridge height and the corresponding ridge widths, for which phase matching is fulfilled (see black crosses in (b)). (d) Effective modal area ($A_{\text{eff}}=(3\cdot\text{Q}_{0000})^{-1}$, see appendix A) of the fundamental $\text{TE}_0$-mode at 780\,nm for the same cladding configurations (i-iii) and related waveguide dimensions as in (c).}
\label{fig:GeoScan}
\end{figure*}

In order to identify the lowest possible pulse energy for optically induced mode conversion we exploit the dependence of the dispersive and nonlinear properties on the geometrical dimensions and cladding configurations of rectangular $\text{Si}_3\text{N}_4$ waveguides. For this type of waveguide there are three possible cladding configurations that are depicted in Fig.~\ref{fig:GeoScan}(a): (i) $\text{Si}_3\text{N}_4$ core buried in $\text{Si}\text{O}_2$ with an air on top cladding, (ii) $\text{Si}_3\text{N}_4$ core with air on top as well as on the sides, and (iii) $\text{Si}_3\text{N}_4$ core surrounded by $\text{Si}\text{O}_2$ on all sides.

The key aspect to achieve efficient mode conversion is, beside fulfilling the phase matching condition, to choose a waveguide with a small amount of group delay per unit length. As the temporal walk-off due to the group delay between the individual control and probe modes reduces the effective interaction length, it also reduces the maximum conversion efficiency at a given control beam power level.
In addition to keeping the effective group delay small, a high conversion rate can partially compensate for the limiting influence of the walk-off on the conversion efficiency. The conversion rate itself increases with the refractive index contrast of the OLPG \cite{Walbaum2013}. Therefore, a high value of the nonlinear coefficient $\gamma$ is needed, which can for instance be accomplished by minimizing the effective modal area.

In order to optimize the three waveguide configurations (i-iii), we first calculate the phase mismatch at the wavelengths studied in this work as a function of the waveguide dimensions. The phase mismatch as a function of the height and width of the ridge for a type (i) waveguide is depicted in Fig.~\ref{fig:GeoScan}(b) exemplarily. Phase matching is not only fulfilled, e.g, for the specific waveguide dimensions displayed in Fig.~\ref{fig:Phase_matching}, but for each displayed ridge height a ridge width can be found, for which the mode conversion process is phase-matched.  The corresponding waveguide dimensions are marked with black crosses in Fig.~\ref{fig:GeoScan}(b). The multitude of phase matching conditions offer the possibility to tailor an optimum waveguide with regard to its dispersive and nonlinear properties, while still fulfilling the phase matching condition.

As a measure of the intermodal walk-off, the standard deviation,
\begin{equation}
\sigma_{\text{GD}} = \left(\frac{1}{n-1} \sum_{i=1}^{n=4} \left| \beta_{1,i} - \bar{\beta_1}\right|^2\right)^{\frac{1}{2}},
\end{equation}
of the group delays per unit length of the two control beam modes ($i=1,2$) and the two probe beam modes ($i=3,4$) is calculated, with $\beta_1$ being the inverse group velocity. A high deviation of either of the four modes from the mean group delay per unit length ($\bar{\beta_1})$ would lead to a reduced interaction length and thereby to a reduced conversion efficiency. The result is shown in Fig.~\ref{fig:GeoScan}(c), where $\sigma_{\text{GD}}$ is depicted by blue crosses and shown as a function of the waveguide's ridge height. The corresponding ridge widths where chosen such that phase-matching is fulfilled (ridge height and width pairs marked with black crosses in Fig.~\ref{fig:GeoScan}(b)). Furthermore, $\sigma_{\text{GD}}$ is displayed for the two other cladding configurations at their respective phase-matched geometries. For each cladding configuration $\sigma_{\text{GD}}$ varies with ridge height and exhibits a local minimum, being lowest with about 1.44\,ps/cm for a type (i) waveguide. The rapid increase of $\sigma_{\text{GD}}$ to more than 5\,ps/cm at smaller ridge heights shows that the occurring group delay has to be taken into account carefully.

In Fig.~\ref{fig:GeoScan}(d) the effective modal area is depicted as a function of the ridge height for the same cladding configurations already discussed above. Again a local minimum is found for each configuration, this time being lowest for the waveguide surrounded by air on three sides (type ii), as here the mean index difference between core and cladding is maximized, and tight light confinement to the core is achieved. 

In order to provide a qualitative understanding of the difference in performance of the waveguides with respect to the conversion efficiency, that will be demonstrated in the next section~\ref{sec:conv}, one can introduce the inverse product of group delay and effective modal area as a figure of merit (FoM):
\begin{equation}
\text{FoM} = \left( \sigma_{\text{GVD}} \cdot A_{\text{eff}} \right)^{-1}.
\end{equation}
For the waveguide of type (i), exhibiting low group delay per unit length, the FoM is calculated to be maximum at a ridge height of 380\,nm and a corresponding ridge width of 949\,nm with a $\text{FoM} =2.8\,\tfrac{\text{cm}}{\text{ps}\cdot \mu\text{m}^2}$~, while the waveguide of type (ii) with the smaller modal area only features a maximum FoM of about $\text{FoM}=2.2\,\tfrac{\text{cm}}{\text{ps}\cdot \mu\text{m}^2}$ at a ridge height of 360\,nm and a width of 960\,nm.

\section{Efficiency of all-optical mode conversion}
\label{sec:conv}

In this section we present the results on all-optical mode conversion in the two waveguide configurations (type i and ii) and with the respective dimensions that promise the most efficient conversion at a given control beam power level: on the one hand the waveguide of type (i) will be discussed, as it exhibits the lowest group delay per unit length at a height of 380\,nm and a width of 949\,nm and therefore allows for a long interaction length of the probe beam with a high contrast index grating. On the other hand the aforementioned waveguide will be compared to a waveguide of type (ii) showing the highest nonlinear coefficient at a height of 360\,nm and a width of 960\,nm and thereby exhibiting the fastest conversion rate at a given power level. We consider in the following two different scenarios: a probe center wavelength of 780\,nm and correspondingly x-polarized TE probe modes as well as a probe center wavelength of 1030\,nm and y-polarized TM probe modes. The control beam is always assumed to be cross-polarized to the probe beam and at the respective phase-matched center wavelength.

\begin{figure}[htb]
\centering
\includegraphics[width=0.6\textwidth]{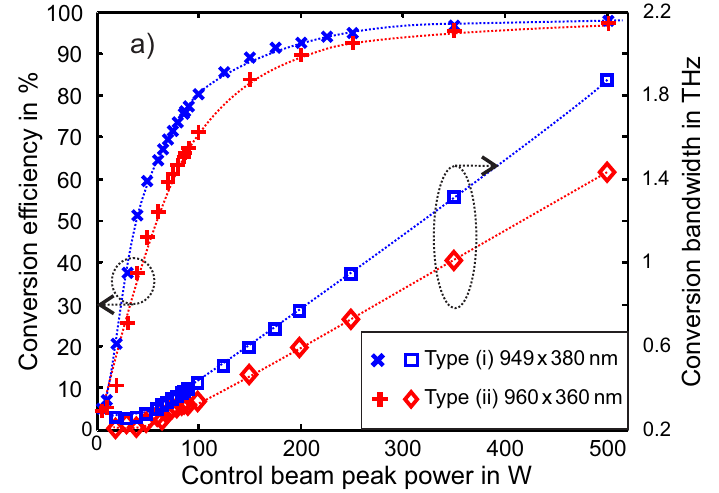}
\includegraphics[width=0.6\textwidth]{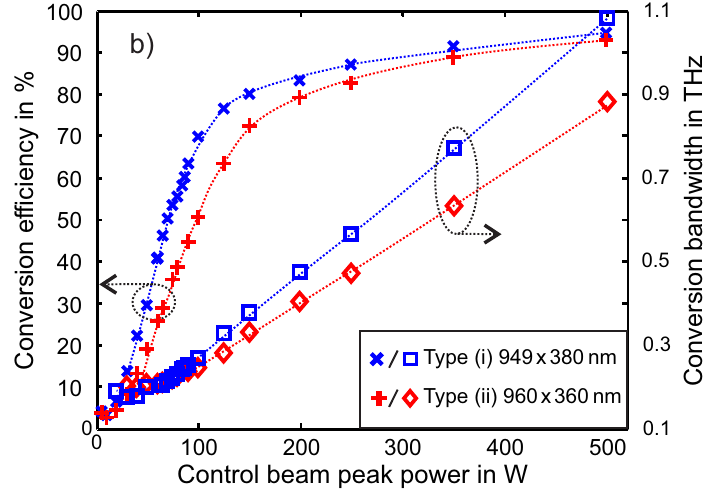}
\caption{All-optical mode conversion efficiency (red pluses and blue crosses) as well as conversion full width half maximum bandwidth (red diamonds and blue squares) as a function of the control beam peak power for two different waveguide dimensions and cladding configurations. The dotted lines are added to guide the eye. The dimensions as well as cladding configurations of the two waveguides are the ones - as discussed in section~\ref{sec:optim} - exhibiting lowest group delay (marked by blue crosses and squares) and highest nonlinear coefficient (marked by red pluses and diamonds), respectively. In (a) the probe pulses are assumed to be in the x-polarized TE-modes and at a center wavelength of 780\,nm while the control beam is assumed to by y-polarized and at a center wavelength of 1030\,nm. In (b) the roles are reversed and the probe beam is assumed to be in the y-polarized TM-modes and at a center wavelength of 1030\,nm and the control beam in the x-polarized TE-modes and at a center wavelength of 780\,nm.}
\label{fig:conversion}
\end{figure}

In order to account for the high differential group delay, pulse durations of 6\,ps for the control beam and 2\,ps for the probe beam were chosen for the simulations. The control beam's energy was evenly distributed between the fundamental and the higher-order mode leading to half the peak power of the whole control beam in each mode. The probe beam energy, being only in the fundamental mode, was set to 1\% of the control beam energy, so that the nonlinear influence of the probe beam on the control beam could be kept negligible. To ensure phase matching and to determine the bandwidth of the conversion process the center wavelength of the beam at about 780\,nm was numerically scanned across the theoretical phase-matched value. The propagation of the involved modes was simulated up to a propagation length of 3\,cm and a periodic energy exchange between both probe modes due to the induced OLPG was observed. The conversion efficiency after the first occurring conversion cycle, i.e., the ratio between probe energy in the higher-order mode and the whole probe energy, is plotted in Fig.~\ref{fig:conversion} as a function of control beam peak power. At each peak power value the conversion efficiency is displayed for probe pulses with a center wavelength phase-matched to the induced grating. In Fig.~\ref{fig:conversion}(a) the mode conversion from the $\text{TE}_0$ to the  $\text{TE}_1$-mode is shown for probe pulses at 780\,nm and a control beam center wavelength of 1030\,nm, while in Fig.~\ref{fig:conversion}(b) conversion from the $\text{TM}_0$ to the $\text{TM}_1$-mode at a center wavelength of 1030\,nm using a control beam at a center wavelength of 780\,nm is displayed.
 
As can be seen, the conversion efficiency increases with control beam energy and thereby with the index contrast of the induced OLPG for both studied waveguides until the conversion efficiency converges towards 100\%. The incomplete conversion at low control beam peak powers despite perfect phase matching can be explained by the low index contrast of the OLPG. The conversion rate scales directly with the grating contrast \cite{Walbaum2013} and at low control beam peak powers the conversion is simply not completed at the waveguide position where control and probe pulses do not overlap in time anymore due to their difference in group velocity. The increasing conversion rate with control beam peak power is then able to compensate at least partially for the limited interaction length and finally leads to a convergence of the conversion efficiency towards 100\% at high peak powers. Comparison of the performance of the two investigated waveguides is revealing that the waveguide of type (i) with the lower intermodal group delay (blue crosses in Fig.~\ref{fig:conversion}) allows for a more efficient conversion at a given peak power, independent of the used probe polarization and wavelength. The $\text{TE}_1$ conversion efficiency exceeds, e.g., 90\% at a control beam peak power of about 150\,W in the waveguide of type (i) while more than 225\,W of control beam peak power is needed in the waveguide of type (ii) to achieve the same efficiency.
This difference in the achieved conversion efficiency is therefore consistent with the introduced figure of merit in section~\ref{sec:optim} being higher for the waveguide of type (i).

 Although the waveguides of type (ii) fall behind in conversion efficiency at a given power level, they might still be of interest from an economical point of view as they do not require a re-coating production step when the waveguides are created by reactive ion etching from a planar $\text{Si}_3\text{N}_4$ layer \cite{Levy2009a}. However, if the available control beam peak power is limited or the relatively new approach of fabricating $\text{Si}_3\text{N}_4$ waveguides by filling predefined trenches in $\text{Si}\text{O}_2$ with $\text{Si}_3\text{N}_4$ \cite{Epping2014b} is used, the waveguides of type (i) are of definite advantage.

 Comparison of the two scenarios displayed in Fig.~\ref{fig:conversion} furthermore reveals, that the conversion seems to be overall more efficient if the probe pulses are centered around 780\,nm and thereby in the x-polarized TE modes and the control pulses are centered around 1030\,nm. The reason for this behavior is twofold: the smaller modal areas at a control beam wavelength of 780\,nm compared to the modal areas at 1030\,nm lead to a stronger self-phase modulation resulting in a rapid broadening of the control beam spectrum. Thereby, the spectral density of the control beam at the phase-matched wavelength decreases more rapidly than for control pulses centered around 1030\,nm leading to a reduced conversion efficiency.  As the 	coupling strength of the grating is a function of the overlap of all involved modes of control as well as probe beam it should not change when switching control and probe beam wavelength.

  However, the numerical model \cite{Poletti2008} used here only includes a frequency dependence of the whole overlap integral describing the nonlinear coupling and not of the individual mode profiles. This constitutes a necessary approximation to be able to compute the simulation within a reasonable time frame. As the coupling coefficient is therefore always evaluated only at the probe frequency when calculating the probe beam propagation, the mode conversion at shorter probe wavelengths than control beam wavelengths is always overemphasized (upper limit) and the mode conversion at longer probe wavelengths than control beam wavelengths is underemphasized (lower limit).

We also investigated the full width half maximum conversion bandwidth for an OLPG, which is defined as the width of the possible probe frequency detuning for which the conversion efficiency exceeds half of its maximum value. The results are displayed in Fig.~\ref{fig:conversion} as a function of control beam peak power. We found a broadening of the conversion bandwidth with increasing control beam peak power, analogous to the rising conversion efficiency with increasing control beam power. Again this can be explained by an increased conversion rate: The fixed dephasing per unit length between the induced grating and the probe modes is compensated by a shorter conversion length leading to a maximum in conversion before a significant dephasing can occur. An analogous behavior can be found for conventional long period gratings, where the achieved conversion does also depend on the phase-mismatch as well as the strength of the grating \cite{Snyder1983}. The bandwidth of the TE conversion process in the waveguide of type (i), e.g., increases from 0.4\,THz as full-width half maximum value at a control beam peak power of 100\,W to 1.9\,THz at a control beam peak power of 500\,W. The slower growth of conversion bandwidth in the waveguide of the second type (ii) is consistent with a steeper slope of the phase matching curve around the phase-matched wavelength of about $13.8\frac{\text{rad}}{\text{cm}\cdot{\text{nm}}}$ compared to $9.6\frac{\text{rad}}{\text{cm}\cdot{\text{nm}}}$  in the waveguide of type (i). In the type (ii) waveguide the accumulated phase mismatch during propagation is larger and, therefore, the conversion bandwidth at a given control beam power is reduced.

\begin{figure}[htb]
\centering
\includegraphics[width=0.8\textwidth]{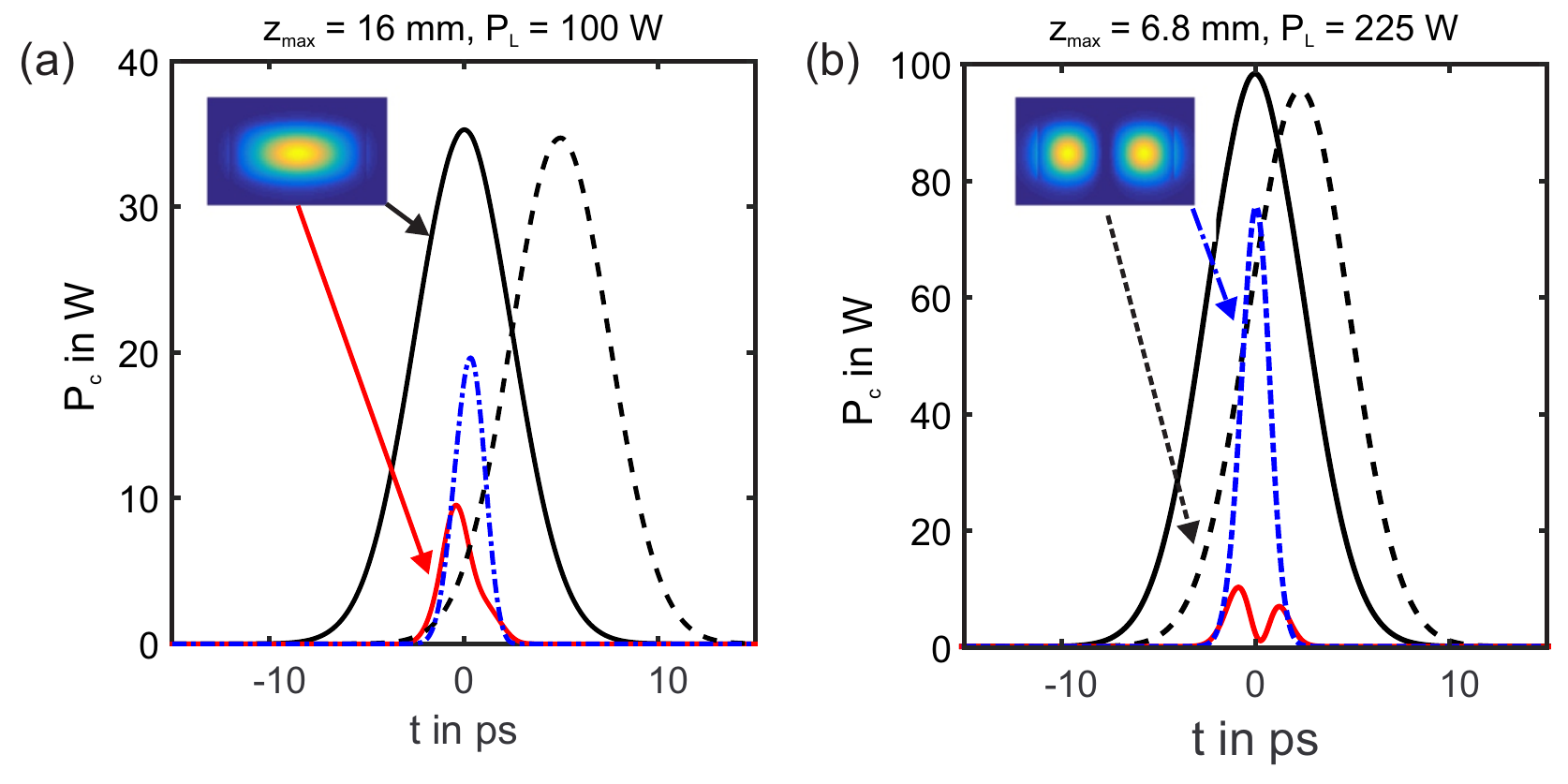}

\caption{Time trace of the individual modes at the waveguide position $\text{z}_\text{max}$, where maximum conversion has occurred in a type (i) waveguide. Displayed are the probe beam pulses at a center wavelength of 1030\,nm (solid red curve: fundamental mode, dash-dotted blue curve: $\text{TM}_1$-mode) as well as the corresponding control beam pulses at a center wavelength of 780\,nm (solid black curve: fundamental mode, dashed black curve: $\text{TE}_1$ - mode). The power of the probe pulses is magnified by a factor of 40 for visibility. (a) The control beam was launched into the waveguide with a peak power of $\text{P}_\text{L}=100\,$W and the inset shows the mode intensity profile of the $\text{TE}_0$-mode, while in (b) the launched peak power of the control beam was 225\,W and the intensity profile of the $\text{TE}_1$-mode is displayed.}
\label{fig:temporal_story}
\end{figure}

In order to visualize the above mentioned influence of the conversion rate in combination with the temporal walk-off on the resulting conversion efficiency the temporal profiles of the involved modes are displayed in Fig.~\ref{fig:temporal_story} exemplarily for two different control beam powers at the waveguide position of maximum conversion $\text{z}_\text{max}$. In Fig.~\ref{fig:temporal_story}(a) the temporal pulse profiles for an initial control beam peak power of 100\,W, resulting in a maximum conversion at a waveguide position of about $\text{z}_\text{max}=16$\,mm, are shown. A significant amount of probe pulse energy is still contained in the fundamental mode. In this particular waveguide configuration the main group delay with about 5\,ps relative to the co-moving time frame is accumulated by the $\text{TE}_1$ control pulse  resulting in a strongly reduced index contrast of the induced grating. When increasing the control beam power and thereby also the conversion rate, as it is displayed in Fig.~\ref{fig:temporal_story}(b), the maximum conversion is already achieved at a waveguide position of 6.8\,mm. Here, the walk-off of the $\text{TE}_1$-mode is still well below the pulse duration, resulting in an almost complete conversion of the fundamental probe mode to the $\text{TM}_1$-mode. 

\section{Conclusion}
In conclusion, we demonstrated, by numerically solving the coupled multi-mode nonlinear Schroedinger equations, the potential for ultrafast all-optical switching via transverse mode conversion in integrated waveguides. We showed that a key to successful mode conversion is tailoring the waveguide dimensions for obtaining a suitable form-induced birefringence such that 
 phase matching at different control and probe wavelengths is achieved. A central advantage of the novel scheme is that it circumvents the experimentally observed cross-talk due to nonlinear polarization rotation when working at the same wavelength for control and probe beam \cite{Hellwig2014a}.
Cladding configurations and waveguide dimensions of $\text{Si}_3\text{N}_4$ waveguides were evaluated systematically to achieve efficient conversion at the lowest possible control beam energy.
Choosing different cladding configurations as well as tailoring the waveguide geometry allow to minimize intermodal group delay as well as to maximize the nonlinear coefficient, even under the constraint of a phase matching condition.  A figure of merit was introduced based on the occurring differential group delay as well as on the effective mode area and the conversion efficiencies of the different waveguides were found to agree with this introduced figure of merit. Finally, an excellent conversion efficiency of more than 90\% is found for control beam peak powers exceeding 150\,W equivalent to only 900\,pJ of pulse energy for pulse durations of a few picoseconds. Integrated multimode waveguides thereby should allow to reduce the needed pulse energies by two to three orders of magnitude in comparison to experimentally shown values \cite{Hellwig2014a} utilizing femtosecond pulses in a step-index fiber (120\,nJ of used control beam energy). The demonstrated scheme thereby poses a promising platform for all-optical switching by modulating the control beam power. 

\section*{Acknowledgment}
We acknowledge support by the Deutsche Forschungsgemeinschaft and the Open Access Publication Fond of the University of Muenster as well as by the Dutch Technology Foundation STW,
which is part of the Netherlands Organisation for Scientific Research (NWO), and which is partly funded by the Ministry of Economic Affairs.

\clearpage

\begin{appendix}

\section{Simulation parameters}
The used parameters for solving the coupled multi-mode nonlinear Schroedinger equations are given in this appendix.
The used nomenclature is from Poletti et al. \cite{Poletti2008} and all values are given in SI units. Of the non-zero values of the coupling coefficients $\text{Q}_{klmn}$ as well as shock time constants $\tau_{klmn}$  (with $k,l,m,n$ being (0,1,2,3) for the ($\text{TE}_0$,$\text{TM}_0$,$\text{TE}_1$,$\text{TM}_1$) modes respectively) only eleven unique values occur that are given in table 1 for their first occurance. The non-zero constants that are not displayed can be derived from the symmetry conditions, e.g., $\text{Q}_{klmn}=\text{Q}_{mnkl}$ for real valued mode fields found in \cite{Poletti2008}. All values are derived at a center frequency of 384\,THz (center wavelength of 780.7\,nm).

\begin{table}[h]
  \centering\caption{Non-zero and unique nonlinear coupling coefficients of the waveguide modes studied in this paper. All parameters are given in SI-units: $\text{Q}_{klmn}$ in $\frac{1}{\text{m}^2}$, $\tau_{klmn}$ in s.}
\begin{tabular}{lcclcc}
 \hline
      & Waveguide (i)        & Waveguide (ii)       &       & Waveguide (i)        & Waveguide (ii)       \\ \hline
$\text{Q}_{0000}$  & 1.4292E+12           & 1.5712E+12           & $\tau_{0000}$ & 5.6381E-16           & 5.2690E-16           \\
$\text{Q}_{0011}$  & 1.3700E+12           & 1.4609E+12           & $\tau_{0011}$ & 6.8667E-16           & 6.7622E-16           \\
$\text{Q}_{0022}$  & 9.4260E+11           & 1.0754E+12           & $\tau_{0022}$ & 6.2714E-16           & 5.4024E-16           \\
$\text{Q}_{0033}$  & 8.5939E+11           & 9.1340E+11           & $\tau_{0033}$ & 7.5969E-16           & 7.2324E-16           \\
$\text{Q}_{0213}$  & 9.1209E+11           & 9.9726E+11           & $\tau_{0213}$ & 7.3787E-16           & 6.8590E-16           \\
$\text{Q}_{1111}$  & 1.3693E+12           & 1.4346E+12           & $\tau_{1111}$ & 7.4827E-16           & 7.4010E-16           \\
$\text{Q}_{1122}$  & 9.6806E+11           & 1.0890E+12           & $\tau_{1122}$ & 7.1602E-16           & 6.4843E-16           \\
$\text{Q}_{1133}$  & 9.2981E+11           & 9.9523E+11           & $\tau_{1133}$ & 7.7811E-16           & 7.2936E-16           \\
$\text{Q}_{2222}$  & 1.4365E+12           & 1.6723E+12           & $\tau_{2222}$ & 6.3948E-16           & 5.3243E-16           \\
$\text{Q}_{2233}$  & 1.3948E+12           & 1.5441E+12           & $\tau_{2233}$ & 7.4881E-16           & 6.7720E-16           \\
$\text{Q}_{3333}$  & 1.4414E+12           & 1.5675E+12           & $\tau_{3333}$ & 7.8023E-16           & 7.0383E-16           \\
      & \multicolumn{1}{l}{} & \multicolumn{1}{l}{} &       & \multicolumn{1}{l}{} & \multicolumn{1}{l}{}
\end{tabular}
\end{table}

\begin{table}[h]
\centering\caption{Dispersion constants of the waveguide modes studied here. All values are given in SI-units: $\beta_n$ in $\frac{\text{s}^n}{\text{m}}$.}

\begin{tabular}{lcccc}
 \hline
       & \multicolumn{2}{c}{Waveguide (i)}  & \multicolumn{2}{c}{Waveguide (ii)}  \\ \hline
			  &   TE0 mode &  TE1 mode &  TE0 mode &  TE1 mode \\ 
$\beta_0$ & 1.475631E+07           & 1.386068E+07           & 1.465939E+07            & 1.363797E+07            \\
$\beta_1$ & 7.139673E-09           & 7.449951E-09           & 7.170060E-09            & 7.617569E-09            \\
$\beta_2$ & 1.483136E-25           & 1.885453E-27           & 1.199630E-25            & -2.424136E-25           \\
$\beta_3$ & 4.921940E-41           & -5.953454E-41          & 9.371723E-41            & 4.036177E-40            \\
$\beta_4$ & 2.902349E-55           & 2.033786E-54           & 2.041110E-55            & 1.866146E-54            \\
$\beta_5$ & -8.433037E-70          & -7.824425E-69          & -7.059796E-70           & -1.110369E-68           \\
\\
&   TM0 mode &  TM1 mode &  TM0 mode &  TM1 mode \\ 

$\beta_0$ & 1.422207E+07           & 1.343765E+07           & 1.410196E+07            & 1.325710E+07            \\
$\beta_1$ & 7.308650E-09           & 7.542772E-09           & 7.329740E-09            & 7.629886E-09            \\
$\beta_2$ & 1.019897E-25           & 3.539029E-26           & 1.150768E-25            & -4.524213E-26           \\
$\beta_3$ & -1.863676E-40          & -4.911103E-40          & -2.455897E-40           & -4.060367E-40            \\
$\beta_4$ & 1.599374E-54           & 5.337803E-54           & 1.758584E-54            & 4.913113E-54            \\
$\beta_5$ & -4.715635E-69          & -2.099060E-68          & -4.999118E-69           & -1.844994E-68          
\end{tabular}
\end{table}
\end{appendix}


\begin{thebibliography}{10}
\newcommand{\enquote}[1]{``#1''}

\bibitem{Boyd1992}
R.~W. Boyd, \emph{{Nonlinear Optics}} (Academic Press, 2008).

\bibitem{Almeida2004}
V.~R. Almeida, C.~A. Barrios, R.~R. Panepucci, and M.~Lipson,
  \enquote{{All-optical control of light on a silicon chip.}} Nature
  \textbf{431}, 1081--1084 (2004).

\bibitem{Tanabe2007}
T.~Tanabe, K.~Nishiguchi, A.~Shinya, E.~Kuramochi, H.~Inokawa, M.~Notomi,
  K.~Yamada, T.~Tsuchizawa, T.~Watanabe, H.~Fukuda, H.~Shinojima, and
  S.~Itabashi, \enquote{{Fast all-optical switching using ion-implanted silicon
  photonic crystal nanocavities},} Appl. Phys. Lett. \textbf{90}, 031115--1--3
  (2007).

\bibitem{Hu2008a}
X.~Hu, P.~Jiang, C.~Ding, H.~Yang, and Q.~Gong, \enquote{{Picosecond and
  low-power all-optical switching based on an organic photonic-bandgap
  microcavity},} Nat. Photonics \textbf{2}, 185--189 (2008).

\bibitem{Nozaki2010}
K.~Nozaki, T.~Tanabe, A.~Shinya, S.~Matsuo, T.~Sato, H.~Taniyama, and
  M.~Notomi, \enquote{{Sub-femtojoule all-optical switching using a
  photonic-crystal nanocavity},} Nat. Photonics \textbf{4}, 477--483 (2010).

\bibitem{Wild2004}
B.~Wild, R.~Ferrini, R.~Houdre´, M.~Mulot, S.~Anand, and C.~J.~M. Smith,
  \enquote{{Temperature tuning of the optical properties of planar photonic
  crystal microcavities},} Appl. Phys. Lett. \textbf{84}, 846--848 (2004).

\bibitem{Nakamura2004}
H.~Nakamura, Y.~Sugimoto, K.~Kanamoto, N.~Ikeda, Y.~Tanaka, Y.~Nakamura,
  S.~Ohkouchi, Y.~Watanabe, K.~Inoue, H.~Ishikawa, and K.~Asakawa,
  \enquote{{Ultra-fast photonic crystal/quantum dot all-optical switch for
  future photonic networks},} Opt. Express \textbf{12}, 6606--6614 (2004).

\bibitem{Koos2009}
C.~Koos, P.~Vorreau, T.~Vallaitis, P.~Dumon, W.~Bogaerts, R.~Baets,
  B.~Esembeson, I.~Biaggio, T.~Michinobu, F.~Diederich, W.~Freude, and
  J.~Leuthold, \enquote{{All-optical high-speed signal processing with
  silicon-organic hybrid slot waveguides},} Nat. Photonics \textbf{3},
  216--219 (2009).

\bibitem{Ding2014}
Y.~Ding, J.~Xu, H.~Ou, and C.~Peucheret, \enquote{{Mode-selective wavelength conversion based on four-wave mixing in a multimode silicon waveguide },} Opt. Express \textbf{22},
  127--135 (2014).


\bibitem{Richardson2013}
D.~J. Richardson, J.~M. Fini, and L.~E. Nelson, \enquote{{Space-division
  multiplexing in optical fibres},} Nat. Photonics \textbf{7}, 354--362
  (2013).
	
	\bibitem{Andermahr2010}
N.~Andermahr and C.~Fallnich, \enquote{{Optically induced long-period fiber
  gratings for guided mode conversion in few-mode fibers},} Opt. Express
  \textbf{18}, 4411--4416 (2010).

\bibitem{Hellwig2014a}
T.~Hellwig, M.~Schnack, T.~Walbaum, S.~Dobner, and C.~Fallnich,
  \enquote{{Experimental realization of femtosecond transverse mode conversion
  using optically induced transient long-period gratings},} Opt. Express
  \textbf{22}, 24951--24958 (2014).

\bibitem{Hellwig2013}
T.~Hellwig, T.~Walbaum, and C.~Fallnich, \enquote{{Optically induced mode
  conversion in graded-index fibers using ultra-short laser pulses},} Appl.
  Phys. B \textbf{112}, 499--505 (2013).


\bibitem{Bauters2011}
J. Bauters, M. Heck, D. John, J. Barton, C. Bruinink, A. Leinse, R. Heideman, D. Blumenthal, and J. Bowers, \enquote{Planar waveguides with less than 0.1 dB/m propagation loss fabricated with wafer bonding,} Opt. Express  \textbf{19}, 24090--24101 (2011).


\bibitem{Ikeda2008}
K.~Ikeda, R.~E. Saperstein, N.~Alic, and Y.~Fainman, \enquote{{Thermal and Kerr
  nonlinear properties of plasma-deposited silicon nitride / silicon dioxide
  waveguides.}} Opt. Express \textbf{16}, 12987--12994 (2008).

\bibitem{Levy2009a}
J.~S. Levy, A.~Gondarenko, M.~A. Foster, A.~C. Turner-Foster, A.~L. Gaeta, and
  M.~Lipson, \enquote{{CMOS-compatible multiple-wavelength oscillator for
  on-chip optical interconnects},} Nat. Photonics \textbf{4}, 37--40 (2009).

\bibitem{Levy2011a}
J.~S. Levy, M.~A. Foster, A.~L. Gaeta, and M.~Lipson, \enquote{{Harmonic
  generation in silicon nitride ring resonators.}} Opt. Express \textbf{19},
  11415--11421 (2011).

\bibitem{Foster2011}
M.~A. Foster, J.~S. Levy, O.~Kuzucu, K.~Saha, M.~Lipson, and A.~L. Gaeta,
  \enquote{{Silicon-based monolithic optical frequency comb source.}} Opt.
  Express \textbf{19}, 14233--14239 (2011).


\bibitem{Epping2013}
J.~P. Epping, M.~Kues, P.~J.~M. van~der Slot, C.~J. Lee, C.~Fallnich, and K.-J.
  Boller, \enquote{{Integrated CARS source based on seeded four-wave mixing in silicon nitride.}} Opt. Express \textbf{21}, 32123--32129 (2013).


\bibitem{Epping2014b}
J.~P. Epping, M. Hoekman, R. Mateman, A. Leinse, R.~G. Heidemann, A. van Rees, P.~J. M. van der Slot, and C.~J. Lee, K.~Boller, \enquote{High confinement, high yield $\text{Si}_3\text{N}_4$ waveguides for nonlinear optical application,} Opt. Express \textbf{23}, 642--648  (2015).

\bibitem{Subramanian2013}
A.~Z.~Subramanian, P.~Neutens, A.~Dhakal, R.~Jansen, T.~Claes, X.~Rottenberg, F.~Peyskens, S.~Selvaraja, P.~Helin, B.~Dubois, K.~Leyssens, S.~Severi, P.~Deshpande, R.~Baets, and P. Van Dorpe, \enquote{{Low-Loss singlemode PECVD silicon nitride photonic wire waveguides for 532--900 nm wavelength window fabricated within a CMOS pilot line},}
 IEEE Photon. J. \textbf{5}, 2202809 (2013).

\bibitem{Ding2013}
Y.~Ding, J.~Xu, F.~D. Ros, and B.~Huang, \enquote{{On-chip two-mode division
  multiplexing using tapered directional coupler-based mode multiplexer and
  demultiplexer},} Opt. Express \textbf{21}, 10376--10382 (2013).

\bibitem{Poletti2008}
F.~Poletti and P.~Horak, \enquote{{Description of ultrashort pulse propagation
  in multimode optical fibers},} J. Opt. Soc. Am. B
  \textbf{25}, 1645--1654 (2008).

\bibitem{Fallahkhair2008}
A.~B. Fallahkhair, K.~S. Li, and T.~E. Murphy, \enquote{{Vector Finite
  Difference Modesolver for Anisotropic Dielectric Waveguides},} J. Lightwave Technol. \textbf{26}, 1423--1431 (2008).

\bibitem{Bures2009}
J.~Bures, \emph{{Guided Optics}} (Wiley/VCH, 2009).

\bibitem{Ramachandran2005}
S.~Ramachandran, \enquote{{Dispersion-tailored few-mode fibers: a versatile
  platform for in-fiber photonic devices},} J. Lightwave Technol.
  \textbf{23}, 3426--3443 (2005).

\bibitem{Walbaum2013}
T.~Walbaum and C.~Fallnich, \enquote{{Theoretical analysis of transverse mode
  conversion using transient long-period gratings induced by ultrashort pulses
  in optical fibers},} Appl. Phys. B \textbf{115}, 225--235 (2013).

\bibitem{Snyder1983}
A.~W.~Snyder and J.~D.~Love, \emph{{Optical Waveguide Theory}} (Springer US, 1983).


\end{thebibliography}
\end{document}